\documentclass[prb, twocolumn, showpacs, amsmath, amssymb]{revtex4} 
\usepackage{multirow}
\usepackage{bbm}
\usepackage{MnSymbol}
\usepackage{bbold}
\usepackage{booktabs}
\usepackage{epsfig}
\usepackage{graphicx}

\newcommand{\ix}[1]{\ensuremath{\text{#1}}} 
\newcommand{\lead}{\ix{lead}} 
\newcommand{\hc}{\ensuremath{\text{H.c.}}} 

\begin{document}

\title{Quench dynamics of correlated quantum dots}

\author{D.M.~Kennes} 
\author{V.~Meden} 

\affiliation{Institut f\"ur Theorie der Statistischen Physik, RWTH Aachen
  University and JARA---Fundamentals of Future Information
  Technology, 52056 Aachen, Germany}

\date{\today}

\begin{abstract}
We study the relaxation dynamics of a quantum dot with local Coulomb correlations 
coupled to two noninteracting leads which are held in grandcanonical equilibrium. 
Only charge degrees of freedom are considered and the dot is described by a 
model which in the scaling limit becomes equivalent to the interacting resonant 
level model. The time evolution of the current and dot 
occupancy resulting out of changes of the dot-lead coupling,
the dots onsite energy, or the charging energy are studied. Abrupt 
and smooth parameter changes as well as setups with and without driving bias
voltage are considered. For biased
dots we investigate  the often studied response  after turning on 
the dot-lead coupling but also the experimentally 
more relevant case in which the voltage is turned on.  
We identify and explain a variety of interesting 
many-body effects and clarify the role of initial correlations. 
\end{abstract}


\maketitle


\section{Introduction}
\label{sec:introduction}

In condensed matter systems at low temperatures two-particle interactions 
lead to a variety of interesting quantum many-body effects which are 
not accessible by perturbative methods.\cite{Mahan90,Bruus04} 
While in equilibrium one already has an elaborate understanding of many 
aspects of correlations in the field of nonequilibrium physics 
this stage is not reached yet. One can distinguish 
between the physics of the possible (nonequilibrium) steady 
state and that of the dynamics. Here we are interested in the time 
evolution resulting out of parameter changes in quantum dot systems
with a few states tunnel coupled to two noninteracting leads. The dot 
electrons interact via the charging energy and the leads are held in 
grandcanonical equilibrium; possibly with two different chemical 
potentials, that is an applied bias voltage. The parameter changes
might be abrupt or smooth on a scale set by the system's inherent time 
scales, with the ultimate limit  of adiabatic changes; abrupt changes 
are often denoted as {\em quenches.} The progress in nanostructuring 
techniques allows to experimentally investigate quantum-dot setups 
with time dependent parameters in a rather controlled way.\cite{Hanson07} 
They are of interest due to their potential for conventional 
as well as quantum information processing. 

Several many-body methods successfully used in equilibrium were recently 
extended allowing to study the dynamics of correlated quantum dots
out of nonequilibrium states; some of those being (mainly) analytical 
\cite{Hackl09,Kehrein10,Pletyukhov10,Schoeller09} others purely 
numerical.\cite{Anders06,Hackl09,Schmitteckert04,Heidrich-Meisner09,Weiss08,Schmidt08} 
Only a few of these techniques have been used to treat Hamiltonians with
explicitely time dependent parameters. We here apply a recently 
developed approximate semi-analytical method  \cite{Kennes12} which is 
based on an extension of the functional renormalization group (RG) 
approach \cite{Metzner12} to Keldysh Green 
functions.\cite{Jakobs07,Gezzi07,Jakobs10b,Karrasch10a,Karrasch10b} 
It allows to study transport 
through quantum dots with a few energy levels, arbitrary time dependence 
of the single-particle parameters as well as 
the two-particle interaction $U$. The approximation of 
Ref.~\onlinecite{Kennes12} is particularly suited for dots which 
only have charge degrees of freedom, but extensions towards spinful 
systems showing the Kondo effect are envisaged. In equilibrium and 
for the bias voltage driven nonequilibrium steady state functional RG based 
schemes which capture aspects of the (spin) Kondo effect have already been 
developed.\cite{Karrasch06,Karrasch08,Jakobs10a} We here restrict ourselves 
to spinless systems and consider the interacting resonant level model (IRLM) 
\cite{Schlottmann80,Bohr07,Borda08,Karrasch10a,Andergassen11} and 
variants of it. 

For small to intermediate $|U|$ our method was shown to provide reliable 
results for the dynamics  of the IRLM on all time scales when applied 
to the standard setup of an initially {\em uncorrelated} and {\em decoupled} dot of arbitrary 
filling and arbitrary level position $\epsilon$, which at time $t=0$ is coupled 
to two grandcanonical leads held at different 
chemical potentials $\mu_{ L}=V/2$ and $\mu_{ R}=-V/2$; $U$ is turned on at  
$t=0$ as well. Here $V$ denotes the bias voltage. We use units with $e=\hbar=c=1$. 
The underlying initial state will be denoted as the {\em standard 
initial state} in the following and the quench protocol (turning on the level-lead coupling) 
as protocol (i). 
The time evolution of the dot occupancy and the current is characterized 
by an exponential decay towards the $U$-dependent steady state value with two 
rates, which for small $|U|$ are related by a factor of two. The decay is modulated 
by oscillations with frequencies given by $\left|\epsilon^{\rm ren} \pm V/2\right|$ 
and power-law corrections,\cite{Andergassen11} where $\epsilon^{\rm ren}$ 
denotes the (renormalized) level position. The functional RG \cite{Kennes12} in 
particular captures the crucial {\em interaction induced renormalization of 
the decay rates.}\cite{Schlottmann80,Bohr07,Borda08,Karrasch10a,Andergassen11}        

Here we study additional protocols for the parameter changes leading to much richer
dynamics. In protocol (ii) we start out in the standard initial state  
with $\epsilon=0$ and let the system evolve into its 
steady state which has dot filling $1/2$. We then quench the onsite energy 
to a large value. For finite bias voltage the resulting exponential 
decay of the occupancy towards its (new) steady-state value 
is repeatedly interrupted by {\em plateau-like structures.} 
They result from the interference of the oscillatory current and the oscillatory 
occupancies 
of the sites next to the central dot one. For $U=0$ the plateaus vanish. 
Protocol (iii) is related to (ii) with the crucial difference that the onsite
energy is raised gradually over an interval $T_{\rm q}$. 
For $V=0$ and not too small $T_{\rm q}$ the occupancy smoothly decreases
and oscillations are damped; the system adiabatically
follows the parameter change. No characteristic
correlation effects appear. This changes for $V>0$. For sufficiently large 
$T_{\rm q}$ we observe an initial {\em increase of the dot occupancy} 
when $\epsilon$ is gradually raised which 
vanishes for $U \to 0$. Studying the renormalization of the dot-lead
couplings we obtain a detailed
understanding of this surprising interaction effect. For both protocols we clarify 
the role of {\em initial correlations} \cite{Stefanucci04} by turning on $U$ not at $t=0$ 
but at the time at which $\epsilon$ is raised. We find that in the present setup correlations
build up rather quickly. In protocol (iv)
we again start in the standard initial state with large $\epsilon$ as well as $\epsilon=0$ 
and $V=0$. After 
the system reached the steady state we abruptly turn on a bias voltage $V>0$. To 
allow for this experimentally more relevant situation (as compared to protocol (i)) 
we have to slightly extend 
the formalism presented in Ref.~\onlinecite{Kennes12}. 
Interestingly, for large $\epsilon$ the initial dynamics of the dot occupancy 
resulting out of this quench differs sizably from the one of protocol (i) with the bias being present 
at $t=0$. This already holds for $U=0$. For $\epsilon=0$ we identify interesting correlation effects.  
The protocols studied here exemplify 
the exceptional flexibility of our approach to the quench dynamics of 
correlated quantum dots. Others can be easily investigated as well. 

Our paper is  organized as follows. In Sect.~\ref{sec:model} we introduce the 
quantum dot model and present the extension of the functional RG approach to
allow for varying the bias voltage with time. Our results for the dynamics
of the different quench protocols are presented in Sect.~\ref{sec:results}. We 
conclude with a brief summary in Sect.~\ref{sec:summary}. 

\section{Model and method}
\label{sec:model}

\subsection{The quantum dot model}

Our model is given by the Hamiltonian 
\begin{equation}
  H(t) = H^\ix{dot}(t) + \sum_{\alpha=L,R} [H^\lead_\alpha + H^\ix{coup}_\alpha(t)] \, .
\label{fullH}
\end{equation}
The dot part $ H^\ix{dot}(t)$ consists of three lattice sites with 
\begin{align}
  H^\ix{dot}_0(t) &= \epsilon(t) n_2 - U(t)\left(\frac{n_1}{2} + n_2 +
    \frac{n_3}{2}\right)   \nonumber
  \\
  & \hspace{6em} + \tau(t)(d_1^\dagger d_2 + d_2^\dagger d_3 + \hc) \, ,
\label{eq:singlepartinter}
  \\
  H^\ix{int}(t) &= U(t) (n_1 n_2 + n_2 n_3),\label{eq:twopartinter}
\end{align}
in standard second quantized notation.
Here $n_j = d_j^\dagger d_j$ denotes the occupancy operator of the spinless 
fermionic level $j$. The levels (sites) are connected locally through a hopping 
matrix element $\tau >0$ and a 
Coulomb interaction $U \geq 0$. Only the central site 2 can be moved in energy by 
changing the `gate voltage' $\epsilon$. The second term in the single-particle part 
of the Hamiltonian is added for mere convenience so that $ \epsilon=0 $ corresponds 
to the point of particle-hole symmetry (half filling of the central site 2 
in equilibrium). 
The two leads $\alpha=L,R$ are modeled as noninteracting,
\begin{equation}
  H^\lead_\alpha = \sum_{k_\alpha} \epsilon_{k_\alpha}
  c^\dagger_{k_\alpha} c_{k_\alpha} \; .  
\label{leadpartH}
\end{equation}
The left lead is tunnel-coupled to side 1 and the right one to site 3 through
\begin{equation}
  H^\ix{coup}_\alpha(t) =  t_{\alpha}(t) \sum_{k_\alpha}
  c^\dagger_{k_\alpha} d_{j_\alpha} + \hc \, ,
\end{equation}
with $j_L=1$ and $j_R=3$. 
We explicitly allow for a time dependence of the parameters 
of $H^\ix{dot}$ and $H^\ix{coup}_\alpha$. We are here interested
in the parameter regime in which band effects do not matter and 
consequently assume the leads to be in the wide band limit
(for details see Ref.~\onlinecite{Kennes12}). The level-lead
coupling is then parameterized by $\Gamma_\alpha(t',t) 
= \pi t_{\alpha}^*(t')t_{\alpha}(t) D_\alpha$, with the constant 
density of states $D_\alpha$ of lead $\alpha$. To reduce the number
of parameters we focus on $\Gamma_L = \Gamma_R = \Gamma$. The leads 
are held at grandcanonical equilibrium with temperature $T=0$ and chemical
potential $\mu_{L/R} = \pm V/2$, with the bias voltage $V>0$. 

In the scaling limit $\tau, |\epsilon|, |U|, V \ll \Gamma$ 
the model is equivalent to the field theoretical  
IRLM \cite{Schlottmann80} with a single energy level as the first and third 
dot sites can effectively be incorporated 
into the leads.\cite{Karrasch10a,Kennes12} Here we do not solely focus on the scaling limit but
also consider setups with $\tau \lessapprox \Gamma$ and $U \lessapprox \Gamma$. 
For $\tau \gtrapprox \Gamma$ our model describes a dot with three levels. The linear 
geometry and the coupling of sites 1 and 3 to the
leads implies a specific coupling of these three energy levels.   

In Ref.~\onlinecite{Kennes12} we gave a detailed account on how to obtain
approximate expressions for the retarded, advanced and Keldysh Green 
functions and the corresponding self-energies for Hamiltonians of the
above type using functional RG. 
From the Green functions and self-energies the occupancies 
$\bar n_j$ of the dot sites and the left and right 
current $J_{\rm \alpha}$ can be computed using standard formulas also 
given in Ref.~\onlinecite{Kennes12}. Consequently we here refrain from 
presenting any technical details on the functional RG and refer 
the interested reader to Ref.~\onlinecite{Kennes12}. Our approximation 
consists in keeping only the lowest order diagrams in the two-particle 
interaction $U$ on the right hand side of the RG flow equations for 
the vertex functions. Solving these differential 
equations however implies resummations of classes of diagrams leading to 
results beyond simple perturbation theory.   
We e.g.~capture the renormalization of the decay rates characterized by a 
power-law behavior in the bare rates with a $U$-dependent exponent\cite{Schlottmann80,Karrasch10a}  
as well as the power-law correction to the exponential decay in time 
of $\bar n_2(t)$ within protocol (i) with an interaction dependent exponent.\cite{Andergassen11,Kennes12}
Both exponents come out correctly to leading order within our approximation
and the method is thus reliable for  $|U| \lessapprox \Gamma$. We emphasize, that 
within the functional RG based scheme we can study the dynamics on arbitrary time scales.
This has to be contrasted to numerical approaches to time-dependent nonequilibrium transport 
through correlated quantum dots with which large times can currently not be reached. A detailed 
comparison of results for the standard protocol (i) to those 
obtained by other approaches can be found in Ref.~\onlinecite{Kennes12}.
For completeness and comparison the physics of this protocol is briefly summarized in 
Sect.~\ref{subsec:proi}.  
Within our approach the two-particle interaction leads to a 
(time dependent) renormalization of the single-particle parameters. To compute 
observables at the end of the RG flow one has to solve a single-particle problem 
with renormalized parameters indicated by the superscript `ren'.

In protocol (iv) the bias voltage is turned on at a time $t_{\rm q}>0$. This type of time
dependence is not directly captured by the formalism of Ref.~\onlinecite{Kennes12} 
which we thus have to supplement. The necessary steps are described in the next 
subsection. 

\subsection{Turning on the voltage}

In Ref.~\onlinecite{Kennes12} we discussed how to handle time dependent couplings to 
the reservoirs via the time dependent hybridization $\Gamma_\alpha(t',t)$. 
We here show that a gauge argument can be used to shift time dependent chemical 
potentials $\mu_\alpha(t)$ into time dependent hybridizations; this way protocol (iv) 
is accessible to the approach of Ref.~\onlinecite{Kennes12} as well.

For structureless wide band limit leads considered here raising the lead filling, and 
thus the chemical potential, is completely equivalent to raising all the 
energy levels of the leads. Therefore, we can replace the lead part of the Hamiltonian 
Eq.~(\ref{leadpartH}) by   
\begin{equation}
  H^{\lead}_\alpha = \sum_{k_\alpha} \left(\epsilon_{k_\alpha}+\mu_\alpha(t)\right)
  c^\dagger_{k_\alpha} c_{k_\alpha} 
\label{eq:addterm}
\end{equation}
and take the leads distribution functions as Fermi functions with zero 
chemical potential. We then apply the gauge transformation
\begin{equation}
 G(t) = e^{i\sum_{\alpha}\sum_{k_\alpha} \Omega_{\alpha}(t)
  c^\dagger_{k_\alpha} c_{k_\alpha} \; },  
\end{equation}
where $\Omega_{\alpha}(t)$ is a suitable gauge field with $\Omega_{\alpha}(0)=0$. 
The transformed Hamiltonian reads
\begin{equation}
 \bar H(t) = G(t)H(t)G^\dagger(t)+i   \dot G(t)G^\dagger(t),
\end{equation}
which shows that for setting $\dot\Omega_{\alpha}(t)=\mu_\alpha(t)$ the additional 
term in Eq.~(\ref{eq:addterm}) originating from the time dependent chemical 
potential cancels with 
$i\dot G(t)G^\dagger(t)=-\sum_{\alpha}\sum_{k_\alpha} \dot\Omega_{\alpha}(t) 
c^\dagger_{k_\alpha} c_{k_\alpha}$. Since  
$\left[H^\ix{coup}_\alpha(t) ,G(t)\right]\neq 0$ the hopping amplitude 
describing the level-lead coupling in $\bar H$ acquires an 
additional phase factor
\begin{equation}
 t_{\alpha}(t) \to  t_{\alpha}(t)e^{i\Omega_{\alpha}(t)} \, ,
\end{equation}
where $\Omega_{\alpha}(t)=\int_{0}^t dt' \mu_\alpha(t')$. 
The transformed Hamiltonian can then be treated as outlined in 
Ref.~\onlinecite{Kennes12}.  

\section{Results for the quench dynamics}
\label{sec:results}

In this section we present and discuss our functional RG results for 
the time evolution of the 
occupancies $\bar n_j$ and the currents $J_\alpha$ induced by 
the quench protocols defined in the introduction. The focus is 
on the correlation effects. For comparison and to gain a deeper understanding 
we present $U=0$ results when necessary. Already now we emphasize that some of 
the interaction effects
cannot directly be linked to the renormalization of the decay rate, which in the
standard setup protocol (i) was found to be the most dominant effect of the 
two-particle interaction.\cite{Bohr07,Borda08,Karrasch10a,Andergassen11} To keep our
presentation self-contained the physics of protocol (i) is briefly 
summarized in the following.  

\subsection{Protocol (i)}
\label{subsec:proi}

\begin{figure}[t]
\centering
\includegraphics[width=0.9\linewidth,clip]{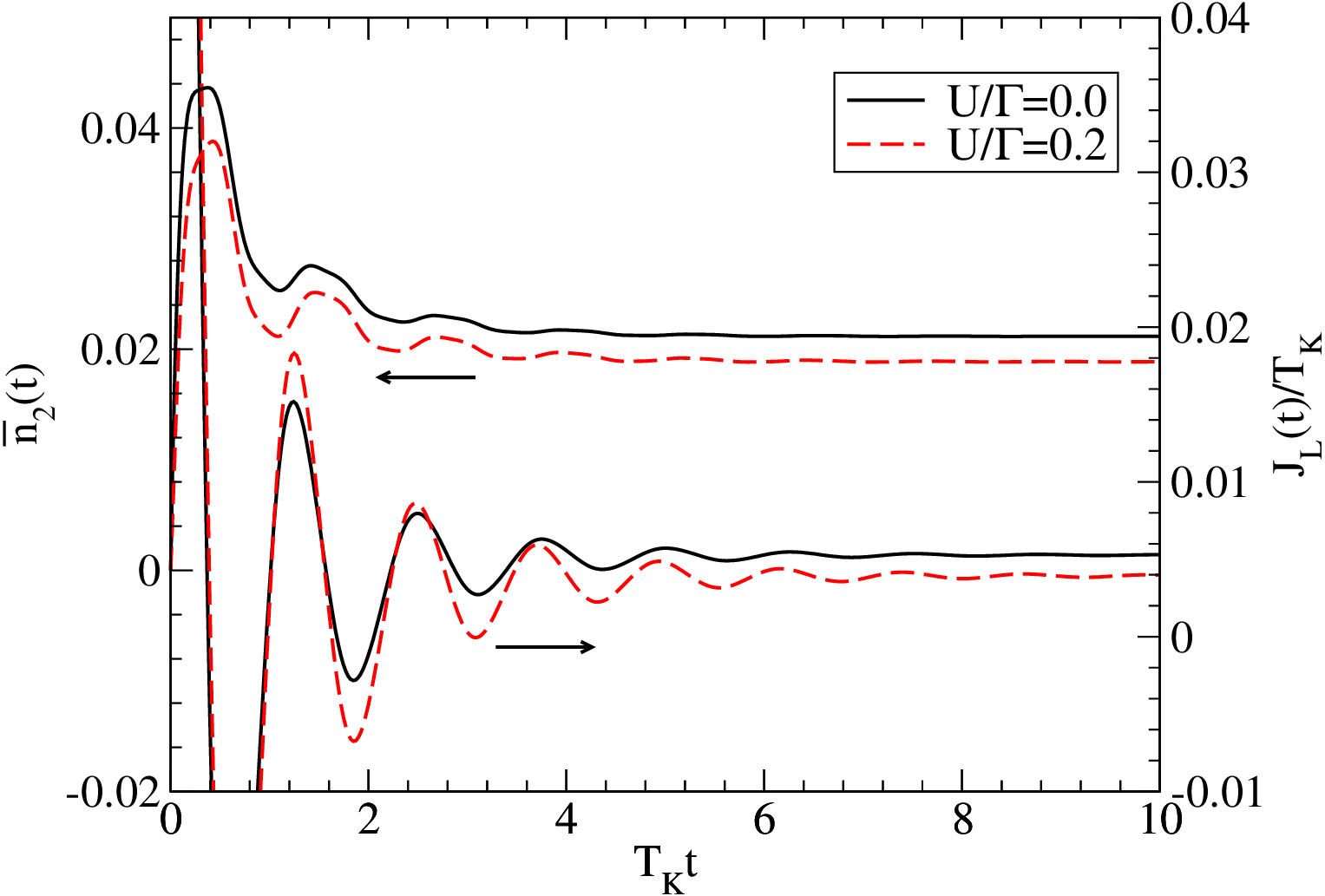}  
 \caption{(Color online) Time evolution of the occupancy $\bar n_2$
and the current leaving the left reservoir $J_L$ for the IRLM in protocol (i). 
The parameters are $\tau/\Gamma=0.0025$, $\epsilon/T_{\rm K}=10$, and $V/T_{\rm K}=10$. 
The universal energy scale $T_{\rm K}$ is renormalized by the local Coulomb 
interaction $U$ (see the text).} 
\label{fig:Timeevolutionocc}
\end{figure}

In protocol (i) we start out with an initially empty dot decoupled from 
the two grandcanonical leads held at chemical potentials $\mu_{L/R} = \pm V/2$.
The coupling to the reservoirs is turned on at time $0$. 
Figure \ref{fig:Timeevolutionocc} shows the time evolution of the level 
occupancy $\bar n_2$ and the current $J_L$ leaving the left lead 
in the scaling limit (the IRLM) for 
$\tau/\Gamma=0.0025$, $\epsilon/T_{\rm K}=10$, and $V/T_{\rm K}=10$. Here $ T_{\rm K} $ is the universal  
energy scale defined by the equilibrium charge 
susceptibility\cite{Schlottmann80,Borda08,Karrasch10a,Andergassen11}
\begin{equation}
\chi=\left.\frac{d\bar n_{2}}{d\epsilon}\right|_{\epsilon=0} = - \frac{2}{\pi T_{\rm K}} \, .
\end{equation} 
For $U=0$ it is given by $4 \tau^2/\Gamma$, but gets strongly renormalized by the 
interaction.\cite{Schlottmann80,Bohr07,Borda08,Karrasch10a,Andergassen11} It increases 
for small to intermediate repulsive interactions. 
The exponential decay of $\bar n_2$ and $J_\alpha$ towards their respective 
$U$-dependent steady-state values is characterized by two relaxation rates. For small 
$U$ they are approximately given by $T_{\rm K}$ and $T_{\rm K}/2$. The decay with the smaller rate is 
accompanied by a power-law correction in $t$ with a $U$-dependent 
exponent.\cite{Karrasch10a,Andergassen11,Kennes12} 
Since this is only a small correction to the exponential decay the main 
effect of the interaction is the strong enhancement of the 
relaxation rates for $U>0$.  For $ |\epsilon^{\rm{ren}}\pm V/2|\gg T_{\rm K},1/t $ 
the decay of the occupancy $\bar n_2(t)$ is overlayed by oscillations with the two frequencies 
$ |\epsilon^{\rm{ren}}\pm V/2|$, where the renormalization of the onsite energy 
$\epsilon \to \epsilon^{\rm ren}$ is of order $(U/\Gamma)^2$ and thus small. In the
current leaving the left (right) lead the frequency $|\epsilon^{\rm{ren}}- V/2|$ 
($ |\epsilon^{\rm{ren}}+ V/2|$) dominates.\cite{Karrasch10a,Andergassen11,Kennes12} 
Comparing the curves for $U=0$ and $U>0$ in Fig.~\ref{fig:Timeevolutionocc} 
it is apparent that the two-particle interaction leads to an increase 
of the oscillation amplitudes. 

\subsection{Protocol (ii)}
\label{subsec:proii}

\begin{figure}[t]
\centering
\includegraphics[width=0.9\linewidth,clip]{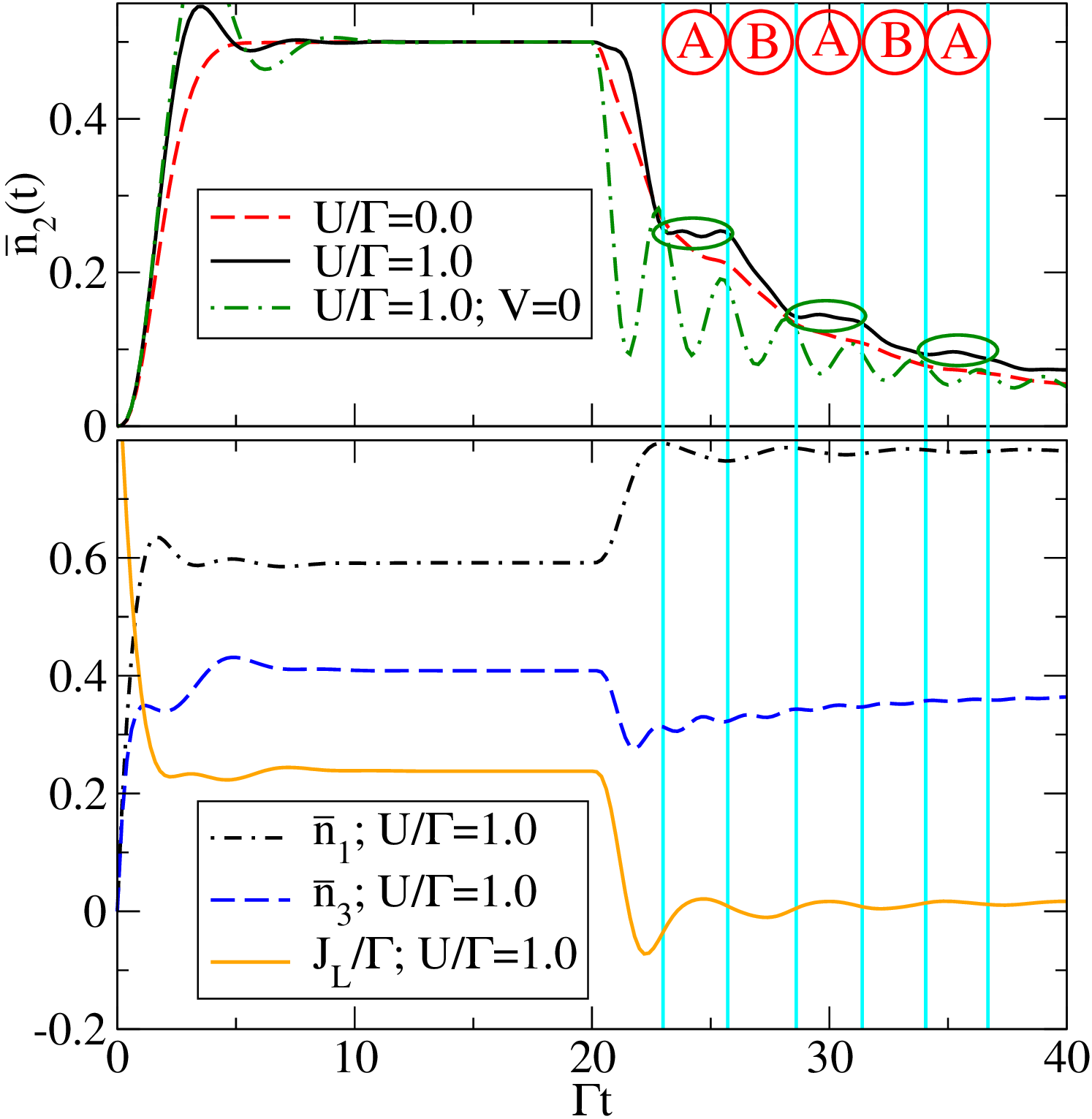}  
 \caption{(Color online) Top: Time evolution of the occupancy $\bar n_2$ 
in protocol (ii). The quench from $\epsilon_{\rm ini}/\Gamma=0$ to $\epsilon_{\rm final}/\Gamma=2$ 
is performed at time $\Gamma t_{\rm q}=20$.
At this time the system has reached a steady state after initially being prepared 
according to protocol (i). 
The other parameters are $\tau/\Gamma=0.5$ and $V/\Gamma=2$. The plateau-like features 
interrupting the exponential 
decay beyond $\Gamma t_{\rm q}=20$ are highlighted by ellipses. For comparison 
$\bar n_2(t)$ for $U/\Gamma=1$ and $V=0$ is shown as well.
Bottom: Time evolution of $\bar n_{1/3}$ and $J_L$.} 
   \label{fig:Quench_Prot_II_a}
\end{figure}

We next study the response of a system, which after being prepared according to protocol (i) 
has reached a steady state, to the sudden increase of $\epsilon$ from 
$\epsilon_{\rm ini}/\Gamma=0$ to the rather large value $\epsilon_{\rm final}/\Gamma=2$ at 
time $t_{\rm q}>0$. We choose $\tau/\Gamma=0.5$, which is no longer deep 
in the scaling limit. The time-evolution 
for $t<t_{\rm q}$ (see Fig.~\ref{fig:Quench_Prot_II_a}) still 
shows the main features found for the IRLM discussed in the last 
subsection. 

Figure \ref{fig:Quench_Prot_II_a} shows the time evolution of the occupancies of  
the three sites as well as $J_L$ for $V/\Gamma=2$. The occupancy of the middle 
site and the current drop quickly after the $\epsilon$-quench at $ \Gamma t_{\rm q}=20 $ 
as the onsite energy is raised far above the chemical 
potential. Whereas in the noninteracting case $\bar n_2(t)$ smoothly decreases 
(up to weakly developed shoulders; dashed line in the upper panel) this depletion is periodically 
interrupted for $U>0$ (solid line; see the ellipses in the upper panel). This effect, which clearly 
goes beyond the renormalization of the decay rates, can 
be understood as follows. 
The interaction enhances the coherence of the system leading to an increase of the 
oscillation amplitudes as  discussed for protocol (i). Plateau-like features show 
up for time regimes where maxima of the current 
leaving the left reservoir coincide with the decreasing part of the oscillatory 
occupancy of the first site $\bar n_1(t)$. 
The combination of these tendencies temporarily stops the depletion process on the 
central site leading to the plateaus. These time regimes are marked with A in 
Fig.~\ref{fig:Quench_Prot_II_a}. For times in which minima 
in the left lead's current coincide with an increase of $\bar n_1$, 
marked with B in Fig.~\ref{fig:Quench_Prot_II_a}, the depletion of the 
central site is even enhanced compared to the noninteracting case. The comparison
to the curve for $U/\Gamma=1$ but $V=0$ (dashed-dotted line in the upper panel) shows
that a finite bias voltage is vital for the development of the plateaus. 

To understand the periodic interception of the decay on a more 
quantitative level we determine the frequencies expected for 
$t>t_{\rm q}$ from protocol (i). For the noninteracting case 
those are given by Eq.~(A.5) of Ref.~\onlinecite{Kennes12} considering 
the poles associated to the smallest relaxation rates (which dominate at large 
times). As a first approximation one can use the same formula with 
the bare parameters replaced by the renormalized ones taken for $t \to \infty$.
This gives an estimate of the corresponding frequencies in the interacting 
case. For the parameters of Fig.~\ref{fig:Quench_Prot_II_a} we obtain
\begin{align}
\omega^{\rm ren}+V/2&\approx 3.23747 \Gamma \\
\omega^{\rm ren}-V/2&\approx 1.23747 \Gamma.
\end{align} 
These two frequencies correspond roughly to the frequencies 
associated to the periodic appearance of the plateaus (smaller frequency) 
as well as the oscillation frequencies within the plateaus (higher frequency).

\subsection{Protocol (iii)}
\label{subsectioniii}

We now focus on the case in which the change of the onsite energy described 
in the last subsection is not abrupt, but is performed with constant velocity 
over the time interval $T_{\rm q}$ centered around $t_{\rm q}$ from 
$\epsilon_{\rm ini}/\Gamma=0$ to $\epsilon_{\rm finial}/\Gamma=2$. To ensure that the 
system has fully reached its steady state at time $t=t_{\rm q}-T_{\rm q}/2$ 
we modify the initial preparation according to protocol (i) 
by starting at $t=0$ with an initial density matrix with occupancy $1/2$ on its diagonal.

\begin{figure}[t]
\centering
\includegraphics[width=0.9\linewidth,clip]{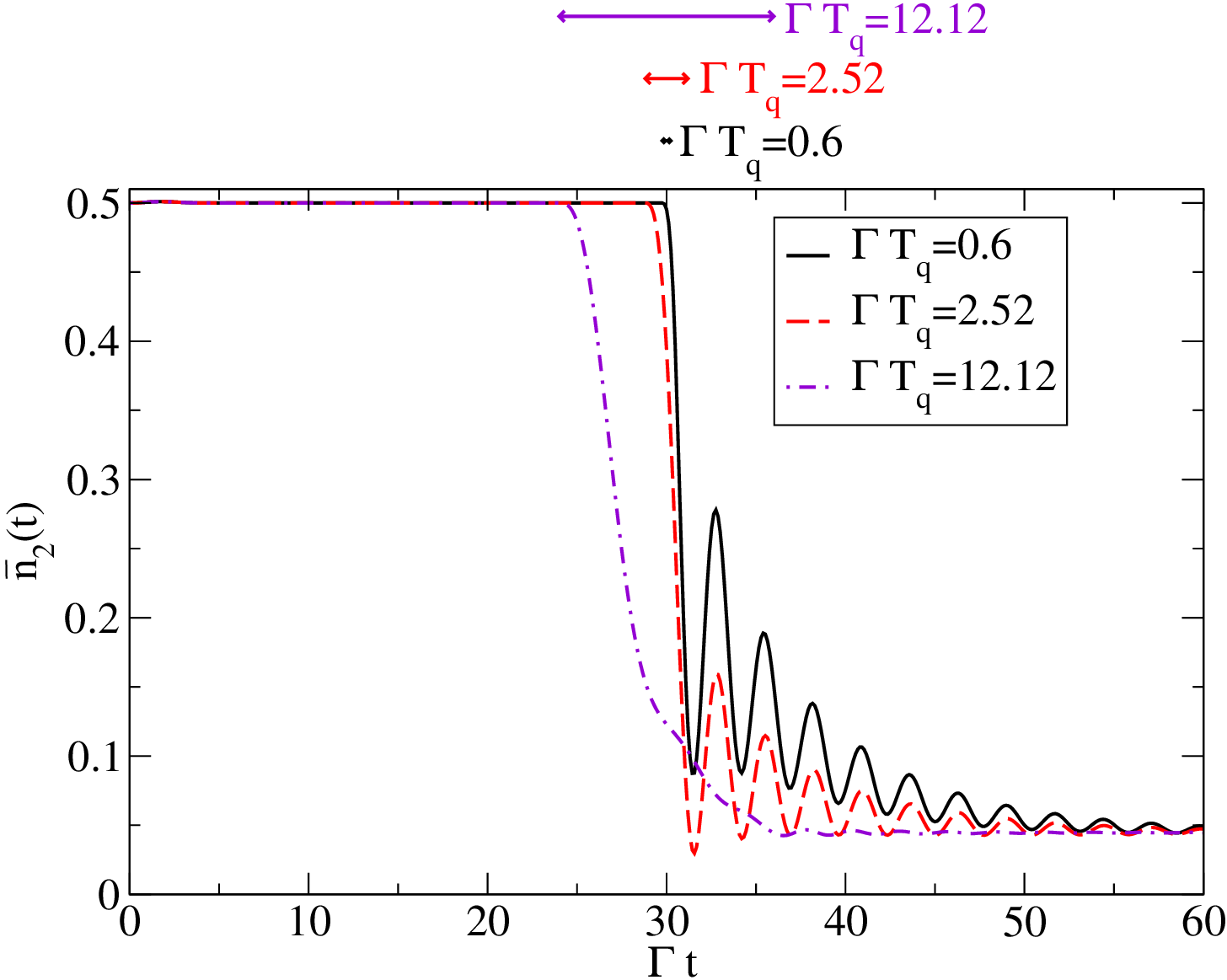}  
 \caption{(Color online) Time evolution of $\bar n_2$ for a 
change of $\epsilon$ performed with constant velocity in the time interval $ T_{\rm q} $
(protocol (iii)). 
The parameter are chosen as in Fig.~\ref{fig:Quench_Prot_II_a} but with $V=0$. 
Arrows above the graph indicate the time interval over which the change of $\epsilon$ 
is performed.} 
   \label{fig:Quench_Prot_III_a}
\end{figure}

We start the discussion with the unbiased case $V=0$ and choose the other 
parameters as $\tau/\Gamma=0.5$ and $U/\Gamma=1$.
Figure \ref{fig:Quench_Prot_III_a} shows the time evolution 
of $\bar n_2$ for different $T_{\rm q}$ (see the arrows above 
the graph) with $\Gamma t_{\rm q}=30$ fixed. As expected the 
amplitude of the oscillations decreases with increasing $T_{\rm q}$. In the 
limit $T_{\rm q} \to \infty$ the system {\em adiabatically} follows the parameter
change and the oscillations disappear.

\begin{figure}[t]
\centering
\includegraphics[width=0.9\linewidth,clip]{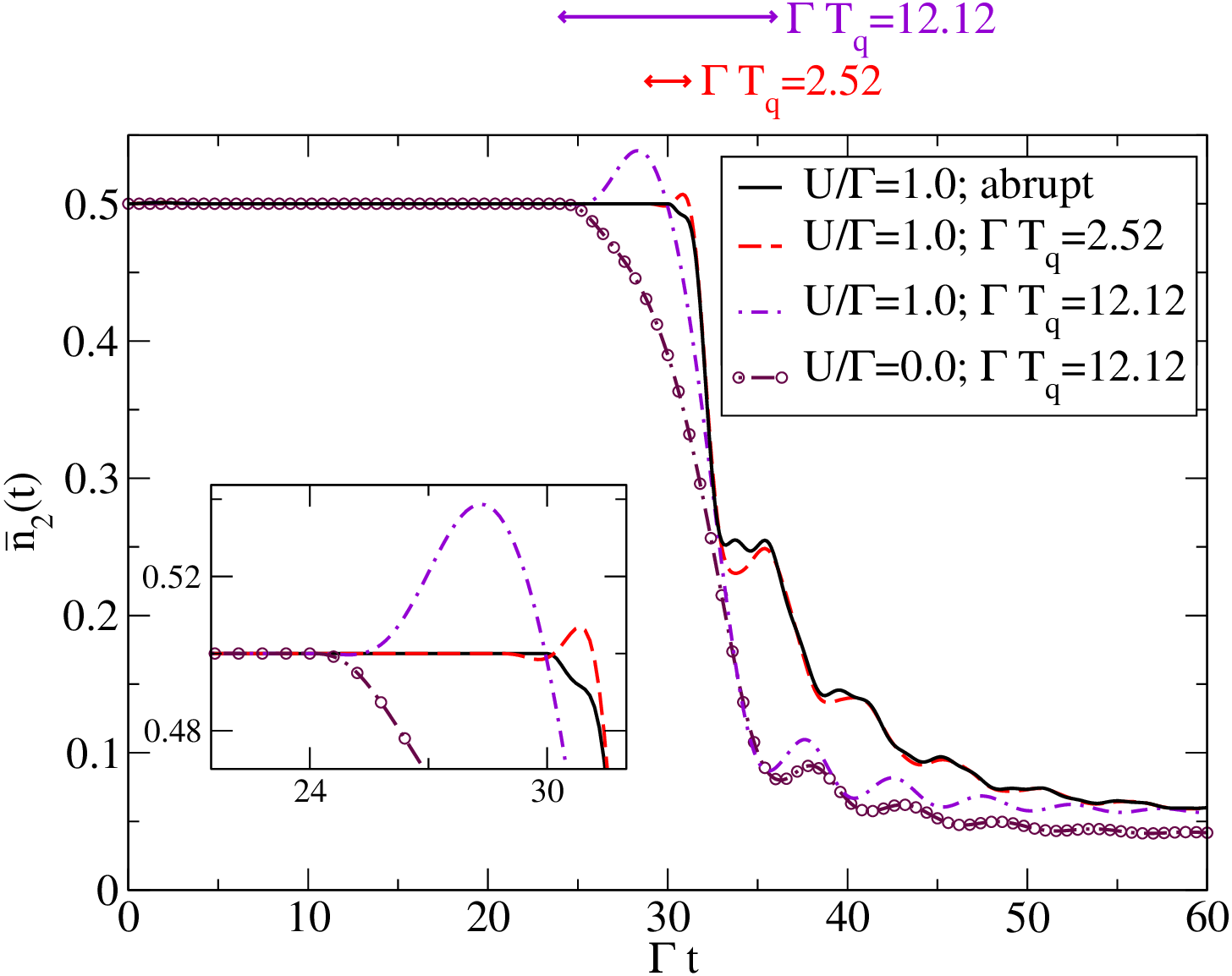}  
 \caption{(Color online) The same as in Fig.~\ref{fig:Quench_Prot_III_a} 
but with $V/\Gamma=2$. For comparison the time evolution after an
abrupt quench (see Fig.~\ref{fig:Quench_Prot_II_a}) is shown as well.  
The inset shows a zoom-in of the region in which $\bar n_2$ (initially) 
increases. Arrows above the graph indicate the time interval over which the 
change of $\epsilon$  
is performed.} 
   \label{fig:Quench_Prot_III_b}
\end{figure}

\begin{figure}[t]
\centering
\includegraphics[width=0.9\linewidth,clip]{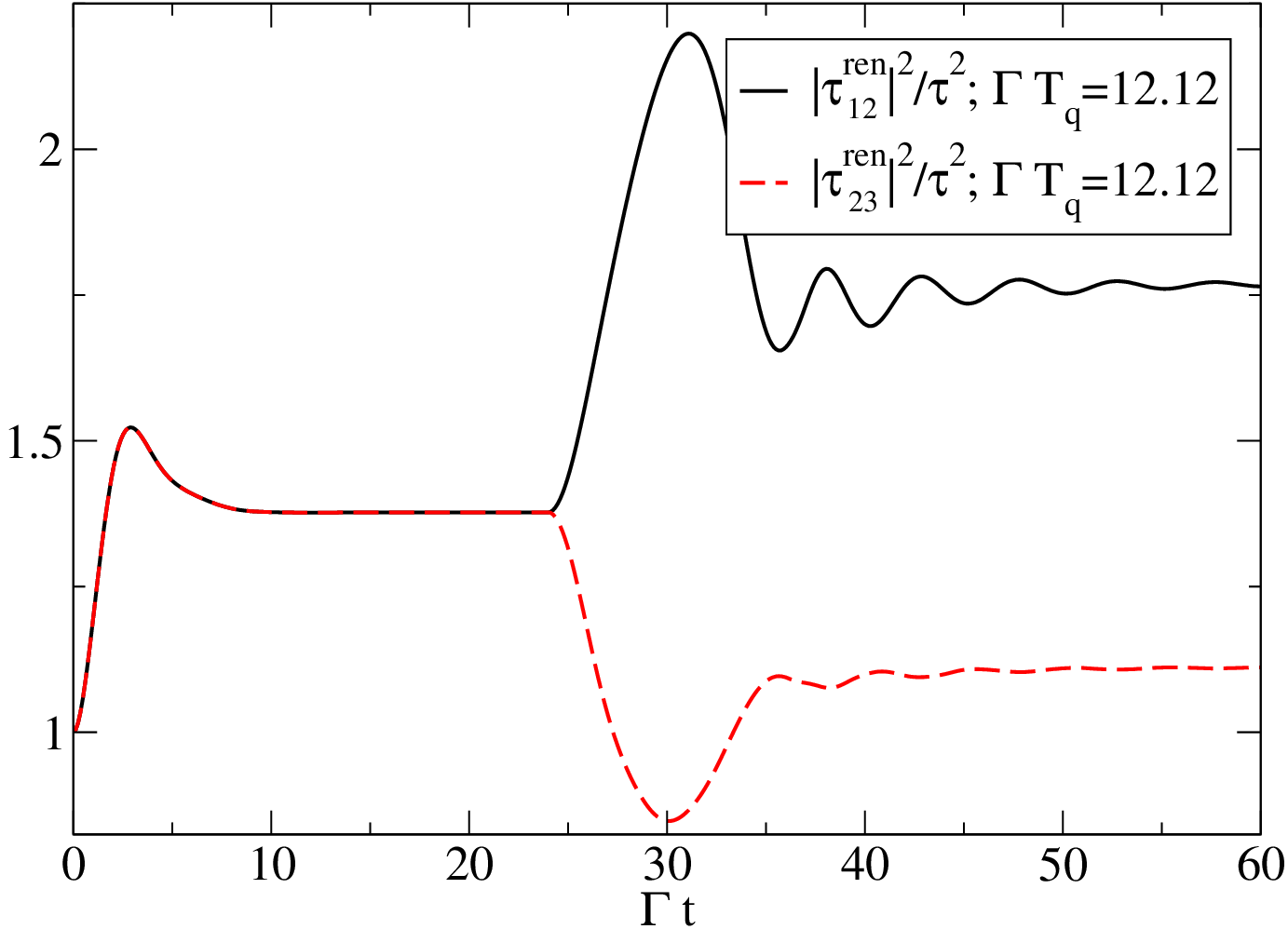}  
 \caption{(Color online) Renormalized hopping amplitudes across bonds  
(1,2) and (2,3) as a function of time (normalized to the bare 
time-independet value). The parameters are chosen as in 
Fig.~\ref{fig:Quench_Prot_III_b} with $\Gamma T_{\rm q}=12.12$.} 
   \label{fig:Quench_Prot_III_c}
\end{figure}

Next we consider the biased case with $V/\Gamma=2$ shown in 
Fig.~\ref{fig:Quench_Prot_III_b}. Interestingly,
for $U>0$ and sufficiently large $T_{\rm q}$, $\bar n_2(t)$ {\em increases}
for times slightly larger than $t_{\rm q}-T_{\rm q}/2$ (see the 
dashed and dashed-dotted lines).
For larger  $T_{\rm q}$ the  effect becomes more pronounced.
There are two competing processes determining 
the occupancy $\bar n_2(t)$. (a) The increase 
of the onsite energy favors a depletion process and dominates in the case 
where the quench is abrupt. (b) At finite bias voltage the interaction
induced renormalization of the hopping 
amplitudes across bonds (1,2) and (2,3) is not equal as shown 
in Fig.~\ref{fig:Quench_Prot_III_c}. The hopping amplitude 
across bond (1,2) (solid line), which is closer to the lead held at higher 
chemical potential, increases more  than the hopping 
across (2,3) (dashed line), by which charge is transported to the lead 
held at lower chemical potential. This induces a tendency towards
an increase of the charge on the central site. 
When the increase of $\epsilon$ is performed slow enough the 
increase of $\bar n_2$ due to the hopping asymmetry 
can initially overcompensate the depletion. At larger times
process (a) wins and the occupancy decreases exponentially 
(with overlayed oscillations). 
We emphasize that the (initial) increase results from the interplay of  
the finite bias voltage $V>0$, the two-particle interaction 
$U>0$, and the parameter change not being abrupt. It vanishes when 
either of these conditions is not fulfilled. 

\subsection{The role of initial correlations}

In protocols (ii) and (iii) the two-particle interaction is turned
on at $t=0$. Correlations are thus fully developed at the time of 
the abrupt or gradual increase of $\epsilon$. To investigate the 
role of these {\em initial} (with respect to $t_{\rm q}$ or 
$T_{\rm q}-t_{\rm q}/2$) correlations in the quench dynamics we 
now modify the `initially correlated' protocols (ii) and (iii) 
and abruptly turn on the interaction at $t_{\rm q}$ or $T_{\rm q}-t_{\rm q}/2$, 
respectively. We will refer to these modified protocols
as `initially uncorrelated'.

\begin{figure}[t]
\centering
\includegraphics[width=0.9\linewidth,clip]{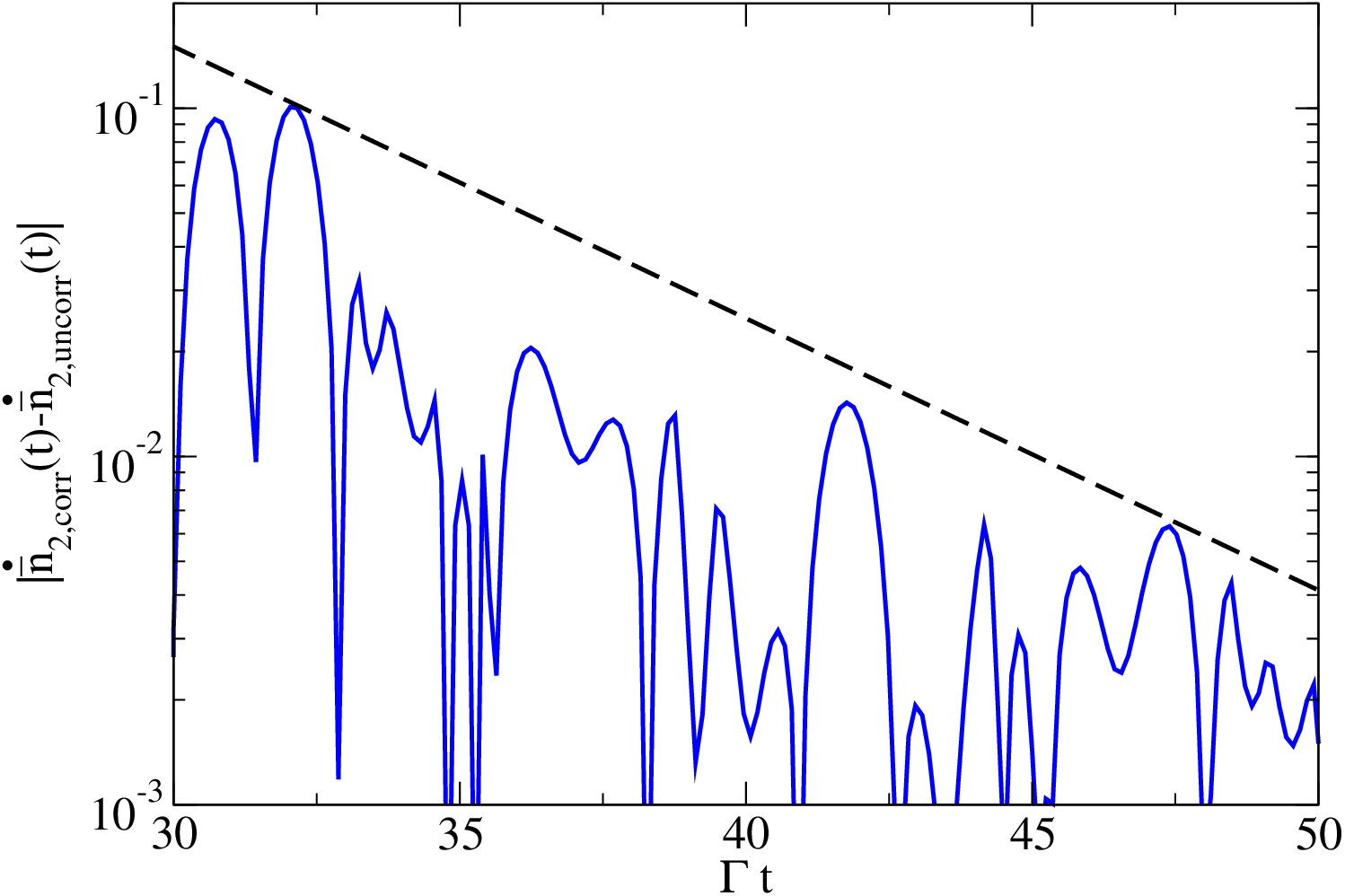}  
 \caption{(Color online) Absolute value of the difference of the derivatives of $\bar n_2(t)$ 
for the initially correlated and uncorrelated protocol (ii). The parameters are
as in Fig.~\ref{fig:Quench_Prot_II_a}. The dashed line displays an exponential
decay with the rate set by the final Hamiltonian. Note the linear-logarithmic scale. } 
   \label{fig:Ini_Corr_a}
\end{figure}

We start out with the comparison of the time evolution of the initially
correlated and uncorrelated protocol (ii). It is not surprising 
that for our tunnel coupled quantum dot setup the 
occupancies $\bar n_2(t)$ of the two situations become equal 
{\em exponentially} fast. In particular,
the plateau-like features discussed in connection with 
Fig.~\ref{fig:Quench_Prot_II_a} are also present in the initially uncorrelated 
setup. To make the
differences clearly visible in  Fig.~\ref{fig:Ini_Corr_a} we show the
absolute value of the derivative of the difference of the two 
occupancies $\bar n_2$ for $t> t_{\rm q}$ on 
a linear-logarithmic scale. The envelope shows a linear behavior which
proofs that the correlations build up on an exponential scale. In fact,
this scale is found to be the decay rate set by the final Hamiltonian (after 
the quench). An exponential decay with this rate is shown as the dashed 
line in Fig.~\ref{fig:Ini_Corr_a}.  

\begin{figure}[t]
\centering
\includegraphics[width=0.9\linewidth,clip]{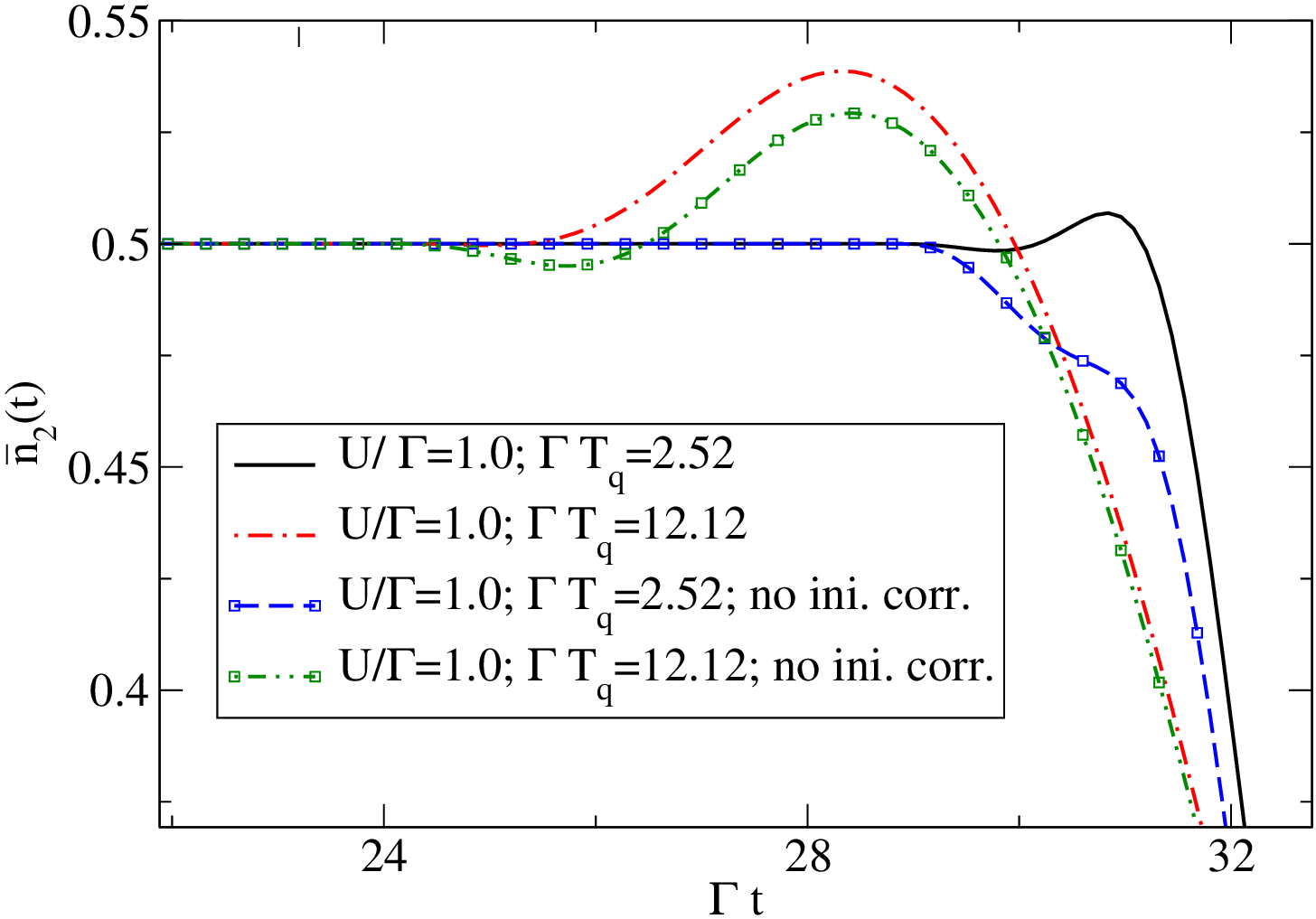}  
 \caption{(Color online) Comparison between the time evolution of $\bar n_2$ 
for protocol (iii) with and without initial correlations. A zoom-in of the interesting
time regime in which $\epsilon$ is raised is shown.} 
   \label{fig:Ini_Corr_b}
\end{figure}

In protocol (iii) we found that the correlation induced 
increase of $\bar n_2(t)$ sets in right after 
the time $t_{\rm q}-T_{\rm q}/2$ at which $\epsilon$ 
starts to be gradually raised. It is thus reasonable
to expect more obvious differences in the dynamics
of the central sites occupancy for the initially correlated
and uncorrelated states of this protocol. As 
Fig.~\ref{fig:Ini_Corr_b} shows the local maximum 
described in subsection \ref{subsectioniii} 
decreases in height if initial correlations are absent (dashed-dotted
line with squares)
up to the extend that the maximum disappears for the faster parameter change  
with $\Gamma T_{\rm q}=2.52$ (dashed line with squares). 
Additionally, the short time behavior right
after the parameter change has been initiated changes significantly. In the
slowest increase of $\epsilon$ with $\Gamma T_{\rm q}=12.12$ (dashed-dotted 
line with squares) first a minimum 
develops. The initial decrease of $\bar n_2$ reflects the 
crucial importance of two-particle correlations for the
accumulation of charge on the central dot. After significant 
correlations have build up (in time) $\bar n_2(t)$ increases 
as found for the initially correlated protocol (iii). The long-time
behavior of the initially correlated and uncorrelated protocols 
(iii) is indistinguishable.

\subsection{Protocol (iv)}

\begin{figure}[t]
\centering
\includegraphics[width=0.9\linewidth,clip]{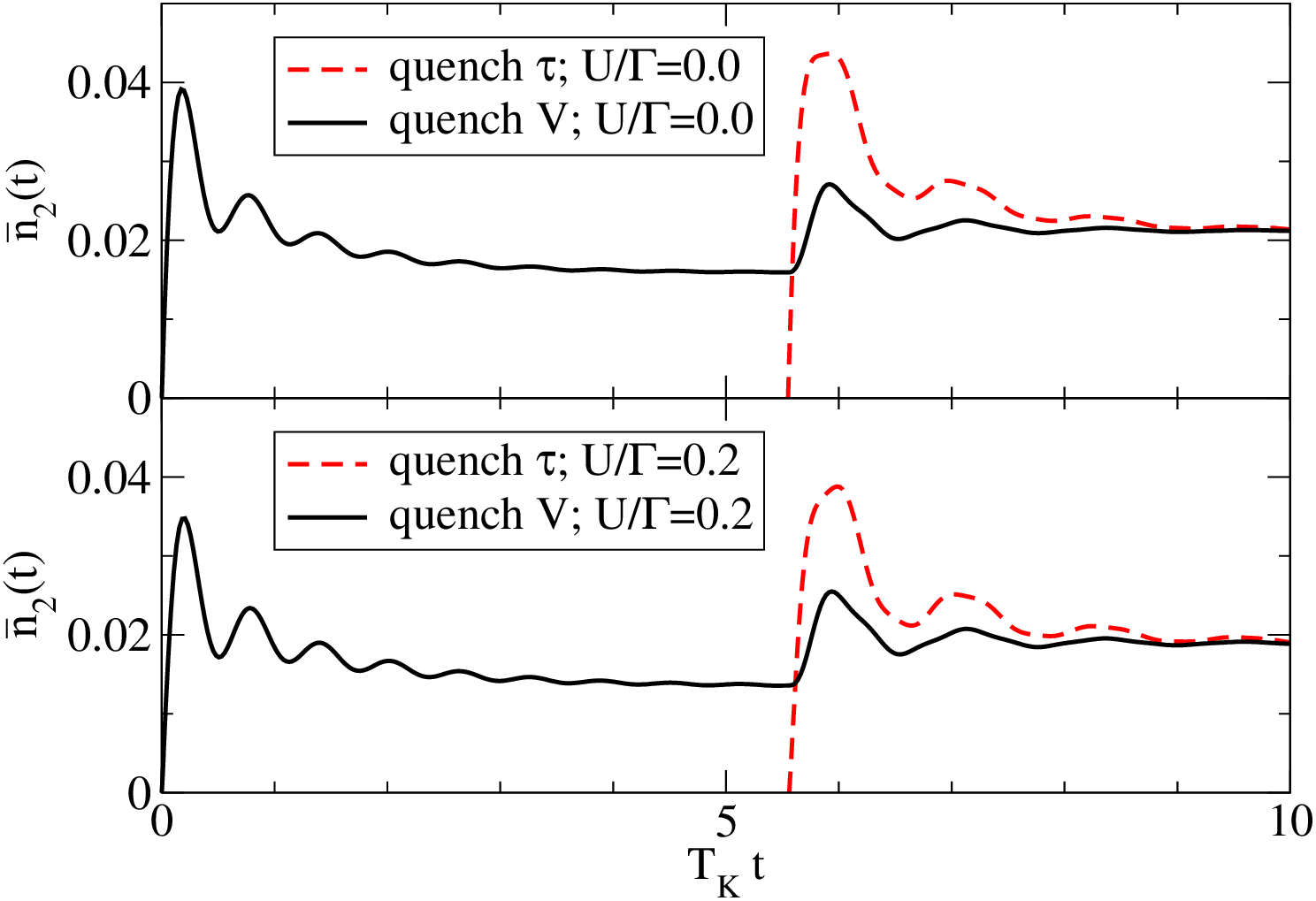}  
 \caption{(Color online) Time evolution of the occupancy 
$\bar n_2$ for a quench in the bias voltage (protocol (iv)) and the 
level-lead coupling (protocol (i)).
The parameters are $V=10T_{\rm K}$, $\tau/\Gamma=0.0025$, 
$\epsilon=10T_{\rm K}$, and $U/\Gamma=0.2$ as well as 0. For a better 
comparison the data or protocol (i) are shifted along the time axis to the 
time of the voltage quench $T_{\rm K} t_{\rm q} \approx 5.5$. } 
   \label{fig:Quench_Prot_IV_a}
\vspace{.5cm}
\centering
 \includegraphics[width=0.9\linewidth,clip]{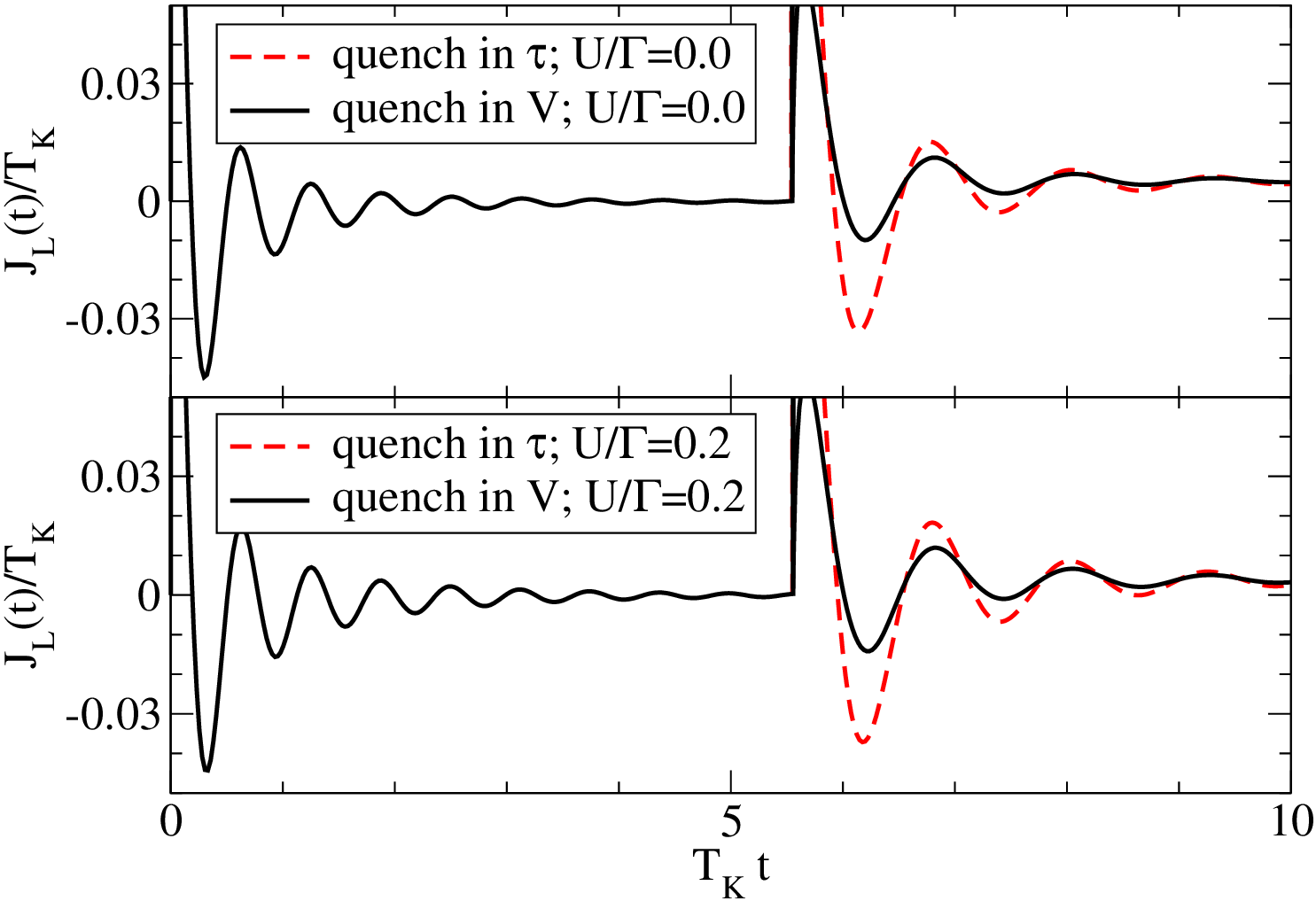}
\caption{(Color online) The same as in Fig.~\ref{fig:Quench_Prot_IV_a} but for 
the current $J_L(t)$.} 
   \label{fig:Quench_Prot_IV_b}
\end{figure}

When studying time-dependent currents and 
level occupancies of correlated quantum dots in most 
theoretical studies the response towards turning on the 
level-lead couplings is considered. In experiments on the other hand one 
generically starts out of an equilibrium state in which the
leads are coupled and at a certain time turns on the 
driving bias voltage. Next we therefore consider this 
experimentally motivated protocol (iv) in the scaling 
limit (the IRLM). For this we start out in the standard initial 
state with $V=0$ and empty dot following protocol (i). After a steady state is reached 
the bias voltage is quenched to a positive value. 
In Fig.~\ref{fig:Quench_Prot_IV_a}  $\bar n_2(t)$ for the abrupt voltage 
quench is compared to the evolution out of the standard 
initial state (with $V>0$) as discussed in 
subsection \ref{subsec:proi}. The parameters are $V=10T_{\rm K}$, 
$\tau/\Gamma=0.0025$,
$\epsilon=10T_{\rm K}$, and $U/\Gamma=0.2$ as well as 0. 
The occupancy for the two cases shows 
sizable differences. While in the evolution out of the
standard initial state the two frequencies 
$|\epsilon^{\rm ren} \pm V/2|$ dominate (see subsection \ref{subsec:proi}) 
the shape is deformed towards a (damped) triangular-like time 
dependence when the bias is only turned on at a later time; instead 
of two, many frequencies $\omega$ with $\epsilon^{\rm ren}-V/2<|\omega|<\epsilon^{\rm ren}+ V/2$ 
contribute to this line shape. The deformation is already
apparent for $U=0$. Furthermore, the amplitude of the oscillations 
is reduced when the bias voltage is quenched. 
In contrast to $\bar n_2(t)$, 
the line shapes of the current $J_L(t)$ in protocols (i) and (iv) shown in 
Fig.~\ref{fig:Quench_Prot_IV_b} are rather similar. The only prominent
difference is the amplitude which decreases if the bias is turned on 
at a later time (also at $U=0$). The correlation effects of 
protocol (iv) for large $\epsilon$  
are the renormalization of the decay rate $T_{\rm K}$ (note that the time axis of 
Figs.~\ref{fig:Quench_Prot_IV_a} and \ref{fig:Quench_Prot_IV_b} is given in units 
of $T_{\rm K}$) and 
the slightly increased amplitude of the oscillation, both known from protocol (i). 

\begin{figure}[t]
\centering
\includegraphics[width=0.9\linewidth,clip]{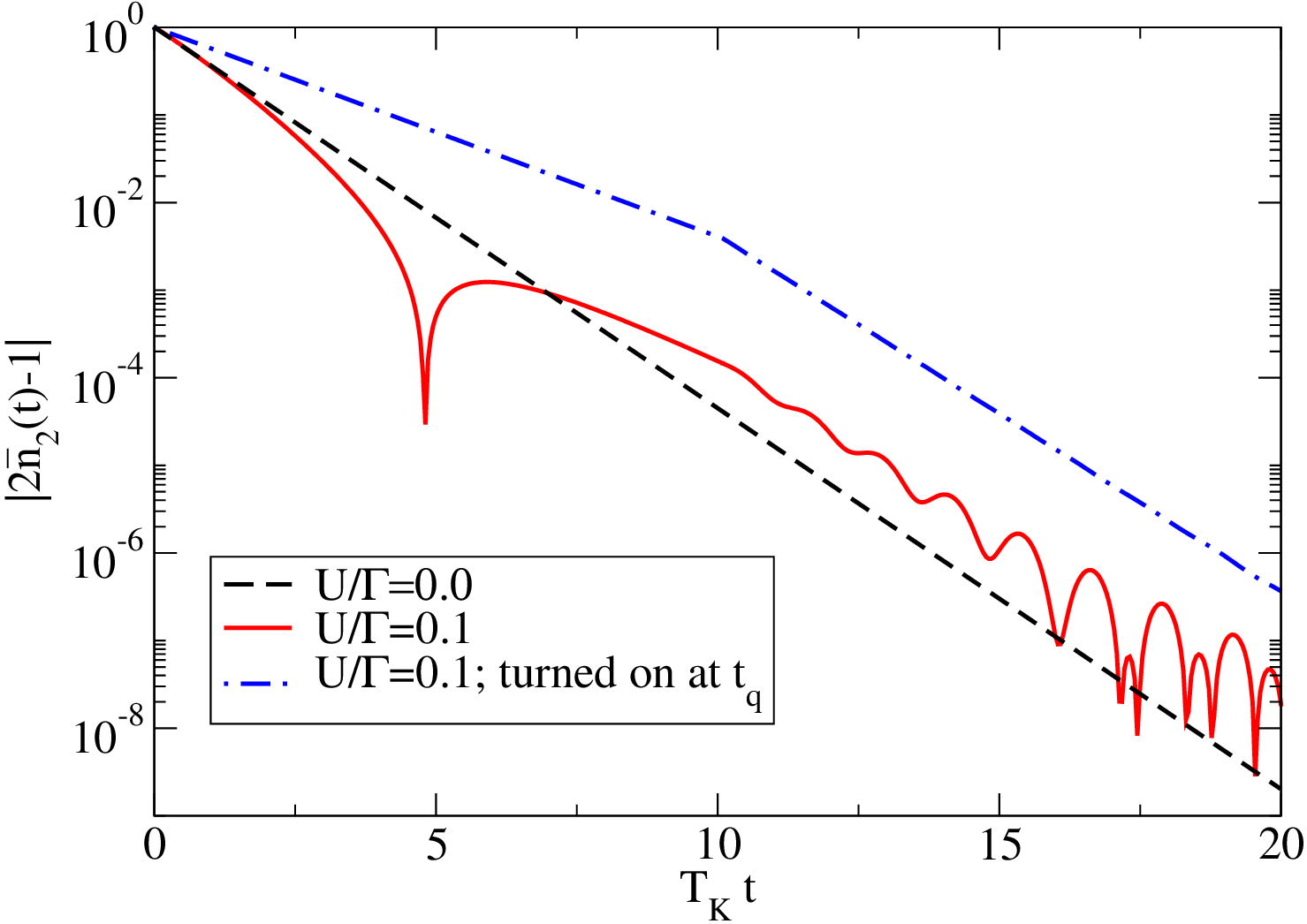}  
 \caption{(Color online) Time evolution of $|2 \bar n_2(t)-1|$ for a quench in the bias 
voltage from $V=0$ to $V=10T_{\rm K}$ at $T_{\rm K}t_{\rm q}=10$. Results for $U/\Gamma=0.1$ 
as well as $0$ are shown. The other parameters are $\tau/\Gamma=0.0025$ and $\epsilon=0.0$. 
For comparison a graph for a setup in which the interaction is turned on simultaneously with 
the bias is shown as well. Note the linear-logarithmic scale. The time axis is given in units 
of $T_{\rm K}(U=0)$ for  the dashed line and $T_{\rm K}(U=0.1 \Gamma)$ for the solid and 
the dashed-dotted line. This leads to the two different slopes and the kink close to 
$t_{\rm q}$ in the dashed-dotted line.} 
   \label{fig:Quench_Prot_IV_c}
\end{figure}

In addition to the case of a large onsite energy we study the dynamics 
for $\epsilon=0$ and initial occupancy $\bar n_2(t=0)=0$. 
The steady-state occupancy equals $1/2$ and it is useful to consider the quantity 
$|2\bar{n}_2(t)-1|$ on a logarithmic scale, which enhances the 
visibility of many features.
For $V=0$ the IRLM is equivalent to the ohmic spin-boson model\cite{Legett87} and 
$|2\bar{n}_2(t)-1|$ corresponds to the absolute 
value of the $z$-component of the spin expectation value.  
Figure \ref{fig:Quench_Prot_IV_c} shows the time evolution 
of $|2 \bar n_2(t)-1|$ out of the standard initial state and a subsequent 
quench in the bias voltage from $V=0$ to $V=10T_{\rm K}$ at $T_{\rm K}t_{\rm q}=10$. 
The noninteracting case as well as $U/\Gamma=0.1$ are depicted. The hopping is $\tau/\Gamma=0.0025$. 
For $U=0$, $|2 \bar n_2(t)-1|$ goes to zero purely exponentially 
with a single dominant rate $T_{\rm K}(U=0)$; straight line (dashed) 
on the linear-logarithmic scale of Fig.~\ref{fig:Quench_Prot_IV_c}.  
This can be understood from the explicit analytical expression 
for $\bar n_2(t)$ in the noninteracting case e.g.~given in  Eq.~(79)
of Ref.~\onlinecite{Kennes12}. For $\epsilon=0$ the two oscillatory terms 
cancel. For $U>0$ one observes the exponentially damped oscillations with
frequency $\sim U$ of the coherent phase of the spin-boson 
model\cite{Egger97,Saleur98,Kennes12} before 
the bias quench (dip for $t< t_{\rm q}$ of the solid line) and additional 
oscillations with much higher frequencies after this. Additionally, 
the dynamics of a setup in which $V$ as well as $U$ are turned on at 
$t_{\rm q}$ (no correlations for $t< t_{\rm q}$) is shown as the 
dashed-dotted line. Although the relaxation rate appears to change to 
its renormalized value almost instantly  as the interaction is turned on
(kink at $t \approx t_{\rm q}$ and the same slope as the other lines for 
$t> t_{\rm q}$; note that the time is given in units of the 
{\em interacting} $T_{\rm K}$ for all times), the dynamics differs 
significantly compared to the one of the solid line; on the scale of the plot
no oscillations are visible. We conclude that in this example the build-up of 
the correlations quickly changes the relaxation rate but does not result in significant 
oscillations as found for the initially correlated case.  
We here concentrated on the occupancy, as the dynamics of $J_L$  
proves less interesting and is characterized by an exponential approach of the steady-state value modulated by 
oscillations with frequency $V/2$.

\section{Summary}
\label{sec:summary}

We have studied the dynamics of the dot occupancy and the current resulting out 
of abrupt quenches and smooth parameter changes in the IRLM and variants of this. 
The time evolution shows characteristics which depend on the protocol with which 
the different parameters are varied in time. In particular, correlations can 
lead to a rich dynamics showing interesting effects which go beyond 
the renormalization of the decay rate found in the standard protocol (i). Using 
our approximate functional RG approach we were able to identify and explain 
several of those. Our method is very flexible, can be used for all time scales,  
and other time dependences of the model parameters than the ones considered here 
can be treated as well. We exemplified the role of initial correlations and 
determined the difference between the dynamics resulting out of switching-on 
the level-lead coupling in the presence of a finite bias voltage and the experimentally 
more relevant case of turning-on the bias voltage in an equilibrium state with 
coupled leads.    

\section*{Acknowledgments}
We thank S.~Jakobs and C.~Karrasch for collaborations in the initial 
state of this work as well as K.~Sch\"onhammer and H.~Schoeller for 
valuable discussions.
This work was supported by the DFG via FOR 723.



\end{document}